\documentclass[preprint,amsmath,amssymb,aps]{revtex4-1}

\usepackage{graphicx}
\usepackage{bm}
\renewcommand{\vec}[1]{\mbox{\boldmath$#1$}}

\begin{document}

\preprint{APS/123-QED}

\title{Magnetic field induced spin wave energy focusing}

\author{Noel Perez} \email{noelpg@usal.es}
\author{Luis Lopez-Diaz}
 
\affiliation{Department of Applied Physics, University of Salamanca, Plaza de los Caidos s/n 37008, Salamanca, Spain}

\date{\today}

\begin{abstract}
	Spin waves can transport both energy and angular momentum over long distances as they propagate. However, due to damping, their amplitude decreases exponentially as they move away from the source, leaving little capability for manipulating how much energy and angular momentum is to be delivered where. Here we show that a suitable local reduction of the effective field can lead to a large accumulation of spin wave energy far from the source. Moreover, both the location and the amount of energy to be delivered can be controlled accurately with geometry and external magnetic fields. Thus, we put forward a general, robust and flexible approach to convey both heat and spin in ferromagnets, which can be directly used in spintronic devices. 
\end{abstract}

\maketitle

The capacity of spin waves to transport both energy and angular momentum as they propagate in a ferromagnetic medium is at the heart of the emerging fields of spin caloritronics \cite{Bauer-12} and magnon spintronics \cite{Kajiwara-10}. To successfully translate fundamental research in these fields to technology, it is crucial to be able to manipulate the flow and delivery of energy carried by spin waves in a controllable and flexible way over different length scales. Since spin wave propagation is mostly governed by intrinsic relaxation mechanisms in the material, strong efforts are being made to develop samples with very low intrinsic damping coefficient \cite{Kelly-13} or to partially compensate damping by means of current-induced torques \cite{Seo-09},\cite{Ando-08} or thermal gradients \cite{Padron-11}. These approaches might lead to a global increase in the propagation length of spin waves, but their decay profile as moving away from the excitation region remains exponential. For energy efficient applications, however, it would be desirable to be able to convey large amplitude spin wave oscillations to specific regions. In a recent paper, An and coworkers \cite{An-13} took a first step in this direction. In particular, they showed that the non-reciprocal nature of magnetostatic surface spin waves (MSSW) allows for controllable unidirectional spin wave propagation and, moreover, for remote heating at the sample edge towards which the spin waves propagate. The mechanism by which the spins pump energy to the lattice leading to remote heating of the sample is, however, not well understood yet.

In the present work we use micromagnetic simulations to demonstrate that a large amount of the energy carried by surface spin waves can be efficiently conveyed to a small area by gradually reducing the effective field at the target region. The mechanism is rather simple. As the field decreases so does the group velocity of the spin waves, and the wave packet starts compressing. At some point, the field reaches a value for which they cannot propagate further, and we observe an increase in the amplitude of the oscillations, which results into localized heating and spin accumulation. We show that the experimental results in [6] can be quantitatively explained within our general scheme considering the decrease in the effective field due to the gradual width reduction at the sample edge. Consequently, we unveil the mechanism by which the spins pump energy to the lattice in \cite{An-13} and, in passing, we propose a general and flexible scheme for focusing the energy flow in a ferromagnet.

\begin{figure}
	\includegraphics[width=1\linewidth]{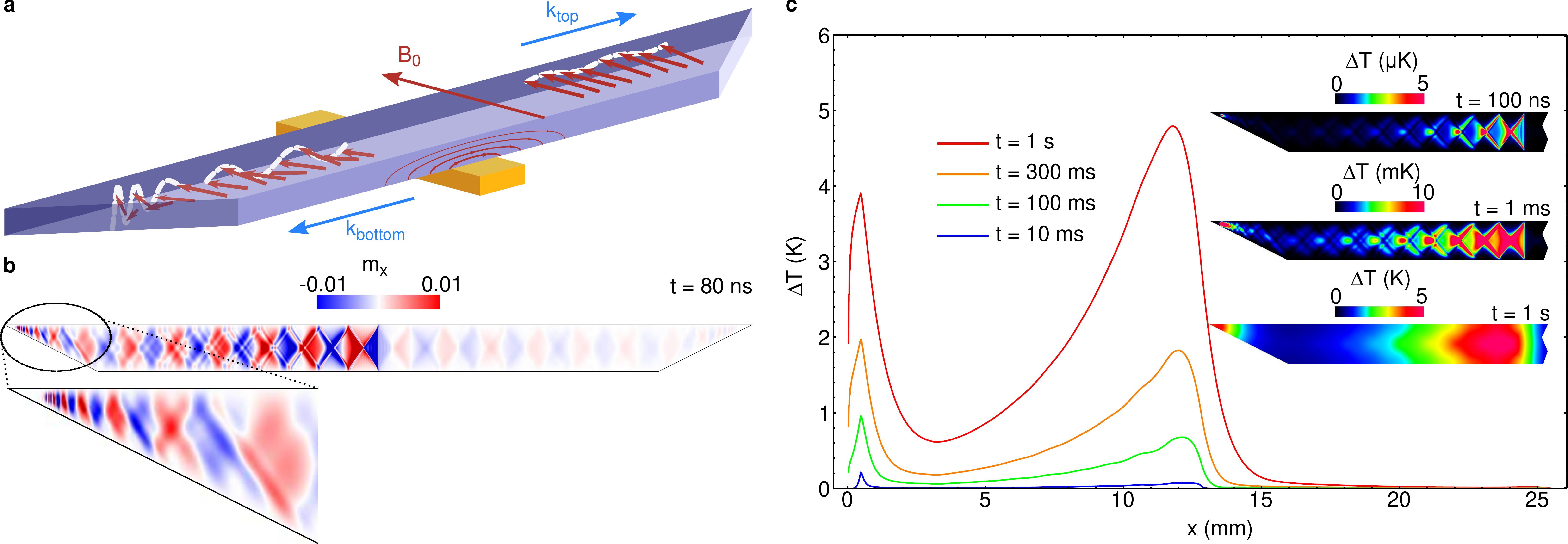}
	\caption{\label{fig_1} \textbf{Geometric control of spin wave energy dissipation. a}, Schematic representation of the studied system. \textbf{b}, Magnetic configuration in the inner layer of the sample ($12 \,\mathrm{\mu m}$ from the bottom) after a time of 80 ns for a continuous microwave excitation at a frequency of 7 GHz. \textbf{b}, Time evolution of the temperature profile of the system. Inset: temperature maps of the propagation region at different times. }
\end{figure}

In this article we study theoretically, for the first time as far as we know, the local variations of the lattice temperature induced by non-uniform magnetization dynamics in the sample. To do that, we have developed a model that couples magnetization dynamics and heat flow in a thermodynamically consistent way (see Methods), allowing us to simulate real-time temperature fluctuations. The system under study is intended to closely mimic one of the experiments in \cite{An-13}. We consider a $25.6 \,\mathrm{mm} \times 1.6 \,\mathrm{mm} \times 32 \,\mathrm{\mu m}$ sample with its edges cut at a sharp angle of $30^{\circ}$ and magnetized in the $y$ direction by a uniform external field $B_0=175 \,\mathrm{mT}$ (see Fig.\ref{fig_1}a). Typical values of single-crystalline YIG film were used for the saturation magnetization ($M_s=1.45\times 10^{5}\,\mathrm{A}\cdot\mathrm{m}^{-1}$), exchange constant ($A=1.0\times 10^{-11}\,\mathrm{J}\cdot\mathrm{m}^{-1}$) and damping constant ($\alpha=5.0\times 10^{-4}$), whereas the thermal behavior was obtained using $\alpha_\mathrm{th}=\frac{\kappa}{\rho\,c_p}=1.5\times 10^{-6}\,\mathrm{m}^2\cdot\mathrm{s}^{-1}$, $\kappa=8\,\mathrm{W\cdot m^{-3}\cdot K^{-1}}$,  $\tau=1.0\,\mathrm{s}$ and $T_0=300\,\mathrm{K}$ for the thermal diffusivity, thermal conductivity, characteristic time constant and room temperature, respectively (see Methods). 

MSSW are excited in the central region with an AC current of amplitude $I_0=18.7 \,\mathrm{mA}$ and frequency $f=7.0 \,\mathrm{GHz}$ along a $50\,\mu\mathrm{m}$ wide antenna placed underneath the film (see Fig.\ref{fig_1}a). Fig.\ref{fig_1}b shows the magnetic configuration of the system after $t=80 \,\mathrm{ns}$. As can be observed, due to the non-reciprocal displacement of MSSW \cite{Stancil-14} and the decay of the AC field amplitude as moving away from the antenna, propagation is mostly unidirectional, since the waves in the bottom surface, moving to the left, are much more intense than the ones in the top surface, which move to the right (Fig.\ref{fig_1}a). Furthermore, the amplitude of the MSSW decays exponentially as moving away from the source due to magnetic damping. At the edge, however, we observe that both the wavelength and group velocity are gradually reduced (see zoomed in region in Fig.\ref{fig_1}(b) and supplementary movie 1), while the amplitude increases. It is worth noting that this energy accumulation occurs before the spin waves reach the leftmost point of the edge, which indicates that pure geometric confinement is not the underlying cause.

After $t\approx200 \,\mathrm{ns}$, the amplitude of the oscillations reaches a stationary distribution and, consequently, so does the heat density distribution delivered to the lattice. Therefore, from then on we evaluate the time evolution of the temperature in the system taking this distribution as an external fixed heat source, which allows us to investigate a time scale in the order of a second, similar to that of the experiments (\cite{An-13},\cite{An-14}). During the first few milliseconds the temperature distribution very closely reproduces the pattern of the MSSW amplitude distribution (Fig.\ref{fig_1}c), but it is smoothed out after a short time due to diffusion, and two temperature peaks become clearly distinguishable, one in the excitation region and the other one at the edge due to the spin wave accumulation described above. In their paper, An and coworkers 
propose two possible mechanisms by which the spin waves release their energy to the lattice, spin wave multi-reflections at the edge due to the sharp $30^{\circ}$ angle cut and two-magnon scattering processes. None of these phenomena were observed in our simulations. Our interpretation, on the other hand, is the following: the reduced width at the edges is responsible for an increase in the demagnetizing field, which in turn reduces the local effective field. The effective field is directly related to the propagation characteristics of the spin waves, and a reduction in its magnitude causes a decrease in the spin waves' wavelength and group velocity, leading to the compression of the wave packet. Eventually, the external field is too low for spin waves to be able to propagate further, and the waves, travelling at almost zero speed, accumulate and dissipate their energy in the form of heat.

\begin{figure}
	\includegraphics[width=0.6\linewidth]{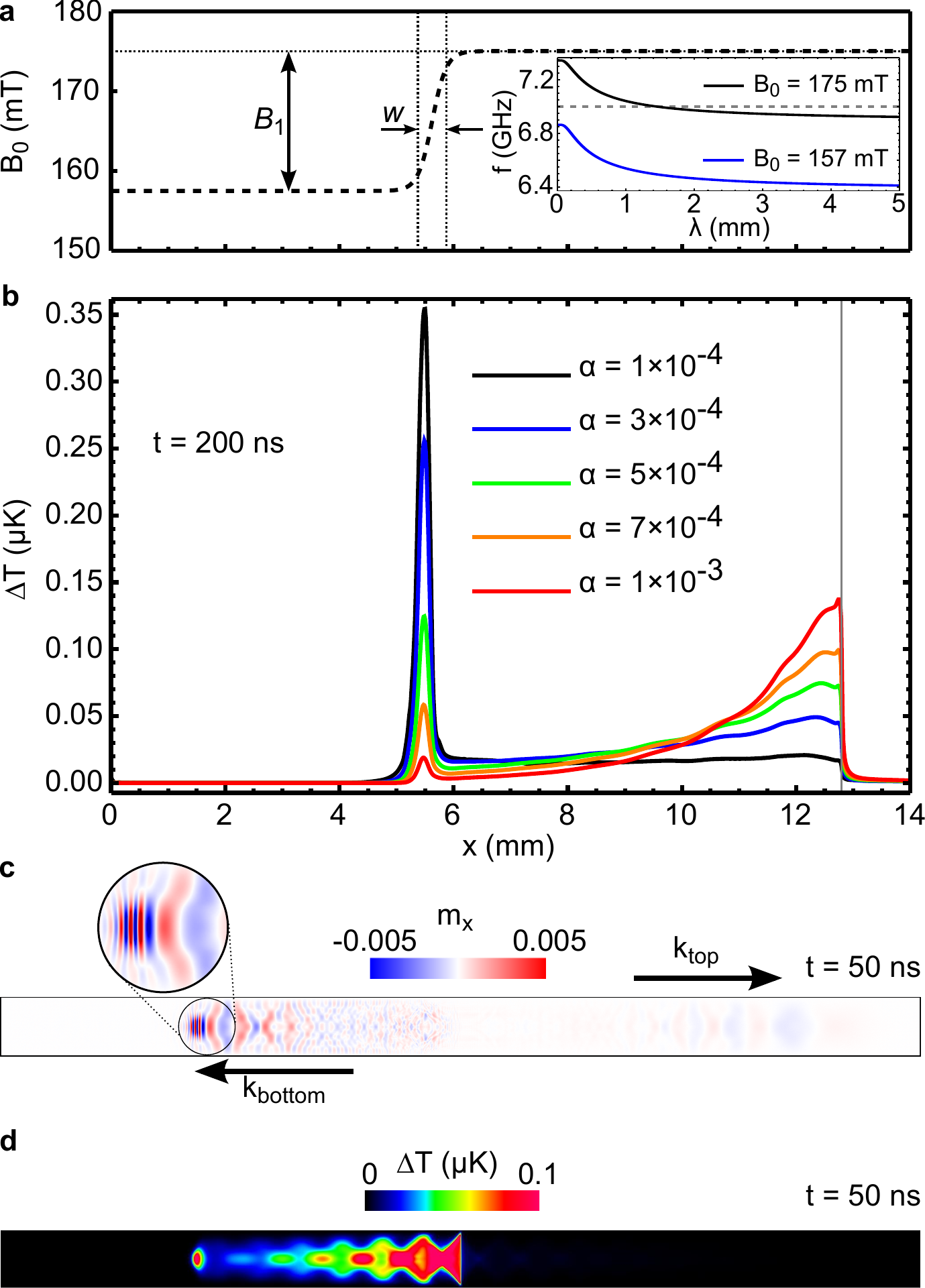}
	\caption{\label{fig_2} \textbf{Spin wave pulse energy focusing by a change in the magnetic field. a}, Spatial distribution profile of the increase of temperature as a function of magnetic damping after a time of 200 ns. External field spatial distribution is represented below. Inset represents surface spin waves dispersion relation in the propagation ($B_0=175\,\mathrm{mT}$) and no propagation ($B_0=157\,\mathrm{mT}$) regions. \textbf{b}, Magnetic configuration after a time of 50 ns. Bottom surface spin waves propagate leftwards whereas top surface spin waves propagate rightwards. \textbf{c}, System temperature distribution after a time of 50 ns.}
\end{figure}
Therefore, the principle underneath spin wave energy dissipation is more general than just geometric confinement; it lies on the propagation characteristics of the waves themselves. MSSW can only propagate in a narrow band of frequencies, and this band can be tuned by controlling the magnetic field acting on the system. If the magnetic field is not uniform and varies in space, so will the properties of the spin waves as they propagate. To further investigate this effect, we perform our next experiment on a system similar to the one described above, but instead of controlling energy dissipation by cutting the edges at $30^{\circ}$ we introduce a spatial variation of the external magnetic field. Namely, we consider a smooth field profile given by $B(x)=B_0-B_1 / (1+\exp\left[\frac{4\,(x-x_0)}{w}\right])$, where $x_0=5.625\,\mathrm{mm}$, $w=0.5\,\mathrm{mm}$ and $B_1=17.5\,\mathrm{mT}$ are the center, width and magnitude of the field variation respectively (see bottom panel of Fig.\ref{fig_2}a). A similar field profile can be experimentally achieved by placing a ferromagnetic stripe at the desired region (see Supplementary figure 1). A $5\,\mathrm{ns}$-long microwave pulse of $7\,\mathrm{GHz}$ frequency is applied through the antenna and the time evolution of the excited spin wave packet is studied.

As in the previous case, the non-reciprocal behavior of MSSW leads to a net transport of energy towards the left. In the excitation region, where $B(x)\approx B_0$, the excitation frequency lies within the range of propagation of MSSW (see inset in bottom panel of Fig.\ref{fig_2}a). When the wave packet reaches the region where the magnetic field changes, the surface waves passband is shifted towards lower frequencies (see bottom panel of Fig.\ref{fig_2}a), and eventually the excitation frequency is left outside the propagation band. At this point, waves cannot propagate further and they are compressed as in the former scenario, leading to localized spin accumulation (Fig.\ref{fig_2}b) and heating (Fig.\ref{fig_2}c). The longitudinal temperature profile at $t=200\,\mathrm{ns}$ is plotted for different values of the damping constant $\alpha$ in Fig.\ref{fig_2}a, all of them within a realistic range for single-crystalline YIG. We observe a high temperature spot localized in the region where the field changes, which becomes more pronounced as $\alpha$ decreases. Therefore, a suitable smooth variation in the external field reveals itself as an efficient way of conveying spin wave energy to a particular region.

\begin{figure}
	\includegraphics[width=\linewidth]{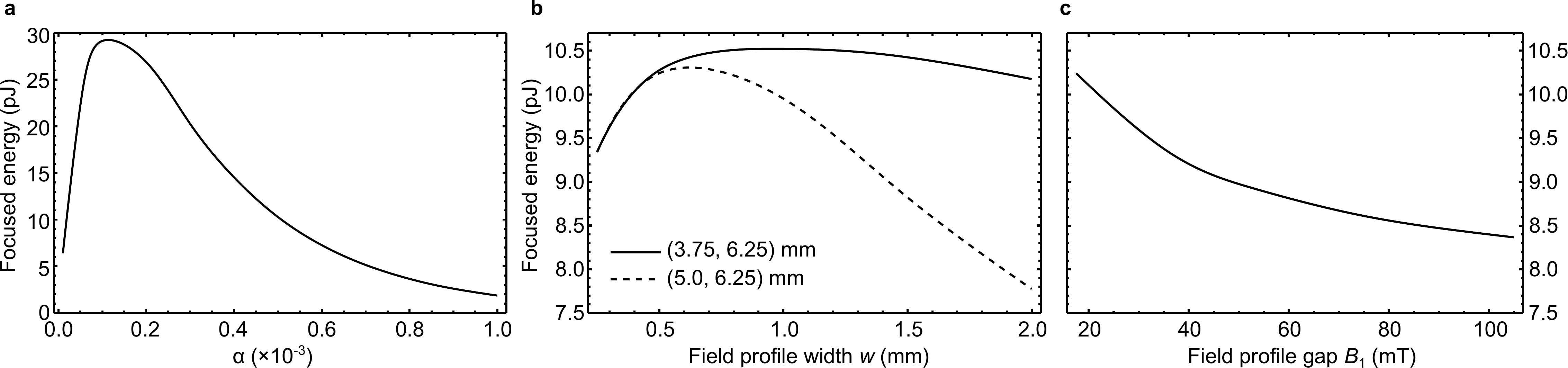}
	\caption{\label{fig_3} \textbf{Energy focusing efficiency with a 5 ns microwave pulse excitation. a}, Energy absorbed at the edge after 200 ns as a function of magnetic damping. \textbf{b}, Energy absorption efficiency as a function of the width of the magnetic field change region ($w$). Solid and dashed lines correspond to energy absorbed in the regions $3.75\,\textrm{mm}<x<6.25\,\textrm{mm}$, and $5.0\,\textrm{mm}<x<6.25\,\textrm{mm}$, respectively. \textbf{c}, Energy absorption efficiency as a function of magnetic field change magnitude. Inset: definition of width and magnitude of the magnetic field jump. In all graphs default parameters are $\alpha = 5\times 10^{-4}$, $w=0.5\,\textrm{mm}$, $B_1=17.5\,\textrm{mm}$, and total pulse energy is 75 pJ.}
\end{figure}

To further explore and optimize this approach for practical applications, we analyze its efficiency as a function of both the damping coefficient $\alpha$ and the magnetic field profile parameters $w$ and $B_1$. As in the previous case, a 5 ns long microwave pulse is applied. The magnetic energy pumped into the magnetic system by the pulse, computed as the energy absorbed in the whole sample over a sufficiently long time, is $E_\mathrm{pulse}\approx 75\,\mathrm{pJ}$. This value was found to be independent of $\alpha$. The focusing efficiency, therefore, is given by the amount of energy dissipated in the region where the field changes. The results are presented in Fig.\ref{fig_3}. Fig.\ref{fig_3}a shows the energy absorbed in the region $5.0\,\mathrm{mm}<x<6.25\,\mathrm{mm}$ after 200 ns as a function of $\alpha$. A well defined maximum at $\alpha\approx 10^{-4}$ is found. Intuitively, a monotonic increase in the efficiency as $\alpha$ decreases would be expected, since lower damping implies less attenuation of the spin waves as moving away from the source. Indeed, this is what we observe for $\alpha > 10^{-4}$. However, a very low damping parameter is also detrimental to the efficiency because a significant amount of energy is reflected back via MSSW in the top surface. 

In Fig.\ref{fig_3}b we show the absorption in both a narrow region ($5.0\,\mathrm{mm}<x<6.25\,\mathrm{mm}$) and a wider region ($3.75\,\mathrm{mm}<x<6.25\,\mathrm{mm}$) as a function of the width of the transition region $w$. The maximum energy dissipated in the narrow region is obtained for $w\approx 0.6$. Below this value both curves coincide, meaning that energy absorption is confined to a small region. For larger values of w, however, where the field variation is more gradual, the two curves diverge, meaning that a significant part of the energy is being absorbed outside the narrow region, resulting in a loss of focus. On the other hand, the efficiency loss for very low values of $w$ observed in Fig.\ref{fig_3}b is attributed to the fact that, for very abrupt transitions, the reflection coefficient increases. The same decrease in efficiency due to reflections is observed if we increase the field variation magnitude $B_1$ instead of decreasing the width, as shown in Fig.\ref{fig_3}c. Consequently, some control of the field variation profile is required in order to maximize the energy delivered to the target region. Nevertheless, the method is robust and has a broad range of operation, since deviations from ideal conditions up to $100\%$ result in only a $10\%$ decrease in the efficiency (Fig.\ref{fig_3}).

The approach for focusing the spin wave intensity with a spatially varying magnetic field proposed in this paper is highly efficient because, as seen in Fig.\ref{fig_3}a, up to $40\%$ of the energy pumped into the system can be conveyed to the target region far away from the source. It is also flexible, since it allows for an easy control of both the target region and the amount of energy to be delivered there. Moreover, we have repeated the experiment discussed in Fig.\ref{fig_2} on a system scaled down by a factor of 100 and the same behaviour was obtained (see Supplementary figure 2), which indicates that the approach is scalable to a large extent. On the other hand, it is also not limited to MSSW, but it can also be applicable to magnetostatic backward volume waves (MBVW, see Supplementary figure 3), although with a significantly reduced efficiency.

Being able to easily control both energy flow and delivery is important not only in spin-caloritronics, but also in spintronic devices, as spin accumulation can also be used to inject a spin current in another ferromagnet or non-magnetic metal adjacent to it. It could also be possible to engineer frequency filters or a magnetic transistor, in which the application of a transversal magnetic field would control whether spin current flows in a waveguide. Its high error-tolerance and down-scalability properties make magnetic field control of spin wave energy an even more promising technique for a wide range of applications. Our method opens a new paradigm for manipulating the spin and energy flow in ferromagnets with external fields. We anticipate a large impact in spintronics, magnonics and spin caloritronics fundamental and applied research.

\section*{Methods}
\label{sec:model}
Our model is meant for ferromagnetic insulators and it assumes that electrons and phonons are in local equilibrium at every moment in time. Consequently, they are considered as a single system referred to as ``the lattice''. The magnetization dynamics is given by the standard Landau-Lifshitz-Gilbert equation

\begin{equation}
\label{eq:LLG}
	\frac{d\vec{m}}{dt}=-\gamma\,\vec{m}\times(\vec{B}_{\mathrm{eff}}+\vec{B}_{\mathrm{th}}) + \alpha\,\vec{m}\times\frac{d\vec{m}}{dt}
\end{equation}

\noindent where $\vec{m}(\vec{r},t)$ is the reduced magnetization, $\gamma=1.76\times 10^{-5} \, \mathrm{T}^{-1}\cdot \mathrm{s}^{-1}$ is the gyromagnetic ratio, $\alpha$ is the damping parameter, $\vec{B}_{\mathrm{eff}}$ is the effective field (including exchange, anisotropy, self-magnetostatic and Zeeman contributions) and $\vec{B}_{\mathrm{th}}$ is the random field representing thermal fluctuations \cite{Palacios-98}. On the other hand, the thermal properties of the lattice are macroscopically described by its temperature distribution, which changes in space and time according to the heat equation

\begin{equation}
\label{eq:heat}
	\frac{dT}{dt}=\frac{1}{c_p \rho} (\kappa \, \nabla^2 T + M_s \frac{d\vec{m}}{dt} \cdot \vec{B}_{\mathrm{eff}} + q_{\mathrm{ext}})
\end{equation}

\noindent where $T$ is the temperature, $\kappa$ the thermal conductivity, $c_p$ the specific heat capacity, $\rho$ the density, $M_s$ the saturation magnetization and $\vec{B}_\mathrm{eff}$ the total effective field acting on the system. The first term accounts for phonon-phonon interactions as a diffusive term, which tends to make the temperature uniform, in a similar way as the exchange does for the spin system. The second term represents heat transfer between the spins and the lattice, computed as the variation of magnetic energy of the spin system. The last term describes heat transfer per unit volume and time between the lattice and the environment. In the present work we consider a standard Newton term $q_\mathrm{ext}=\frac{T_0-T}{\tau}$, $T_0$ being the room temperature and $\tau$ the characteristic thermal relaxation time constant. Equations (\ref{eq:LLG}) and (\ref{eq:heat}) are solved self-consistently, the temperature distribution entering (\ref{eq:LLG}) in the amplitude of the thermal field \cite{Palacios-98}. To solve both equations numerically we discretize the sample in $12.5\,\mu\mathrm{m} \times 12.5\,\mu\mathrm{m} \times 8.0\,\mu\mathrm{m}$ cells. The use of such large cells as compared to the exchange length ($\sqrt{\frac{2\,A}{\mu_0\,M_s}}=27.5\,\mathrm{nm}$) is justified because in all the simulations performed the amplitude of the oscillations is below $1.5$ degrees even in the excitation region.

\section*{Acknowledgements}
This work was supported by the regional government (project SA163A12 from Junta de Castilla y Leon), Spanish government (project MAT2011-28532-C03-01) and the European Social Fund.

\section*{Author contributions}
Both authors contributed equally to the work presented in this paper.

\section*{Competing financial interests}
The authors declare no competing financial interests.
\newpage
\section*{Supplementary material}

\section{Construction of the field profile}

In the main text we impose an external field variation with a profile given by
\begin{equation}
B(x)=B_0-\frac{B_1}{\exp(\frac{4(x-x_0)}{w})+1}
\end{equation}
Our goal was to create two separate regions, one ($x>x_0$) in which surface spin waves propagate and another one ($x<x_0$) in which they do not, with a smooth transition between them. A simple way to achieve a local reduction in the effective field is to place a small magnetic stripe on top of the magnetic waveguide. Both the waveguide and the stripe would be magnetized in the same direction ($y$) by the external field, and the magnetostatic field from the small stripe would act against the external field in the magnetic waveguide (see Fig. S1a). 

We simulated the effective field created by this configuration. The small magnetic stripe has dimensions $3.2 \,\mathrm{mm} \times 1.6 \,\mathrm{mm} \times 24 \,\mathrm{\mu m}$, and its center is situated at $4.8 \,\mathrm{mm}$ from the left edge of the waveguide. The separation between the stripe and the waveguide is $8 \,\mathrm{\mu m}$, and the saturation magnetization in the waveguide is $M_S=1.1\times 10^6 \,\mathrm{A/m}$. Solid line in Fig. S1b shows the effective field profile at the center of the magnetic waveguide for the proposed geometry and a uniform external field $B_0 = 175\mathrm{mT}$.  The dashed line in Fig. S1b shows the effective field profile for a non-uniform external field following equation S1, with parameters $c=6.4\,\mathrm{mm}$, $w=1.6\,\mathrm{mm}$ and $B_1=12\,\mathrm{mT}$.
\begin{figure}[h]
	\includegraphics[width = \linewidth]{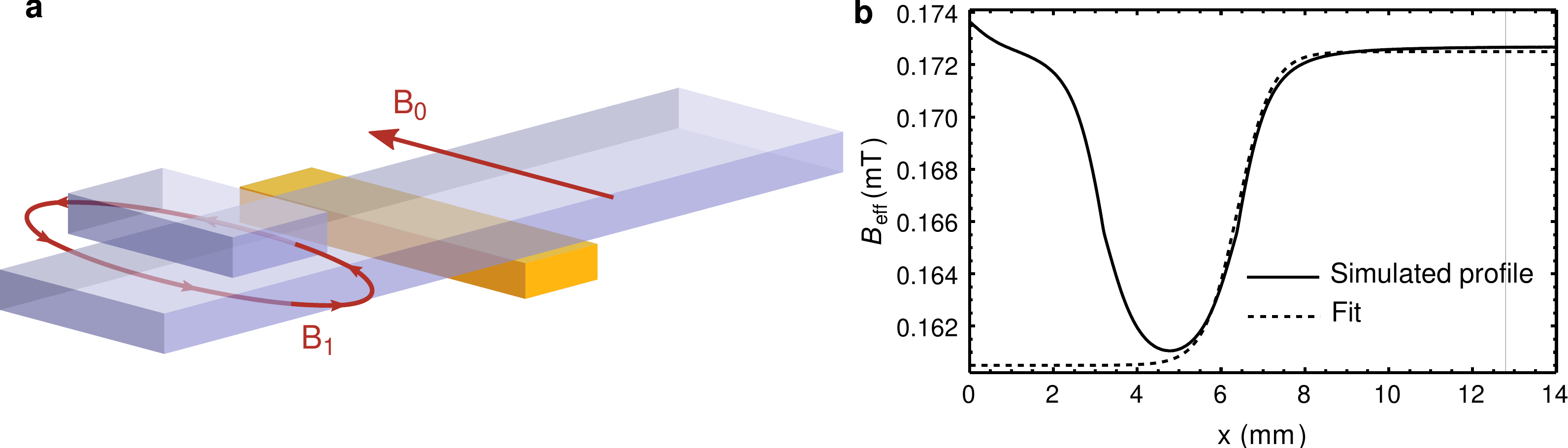}
\end{figure}

\textbf{Figure S1: Experimental realization of a non-uniform external field. a}, Schematics of the proposed geometry. \textbf{b}, Effective field $B_{\mathrm{eff}}$ profile for the proposed geometry (solid line) and fit to a smooth profile (dashed line), for an external field $B_0 = 175\mathrm{mT}$.

\begin{figure}[h!]
	\includegraphics[width=0.6\linewidth]{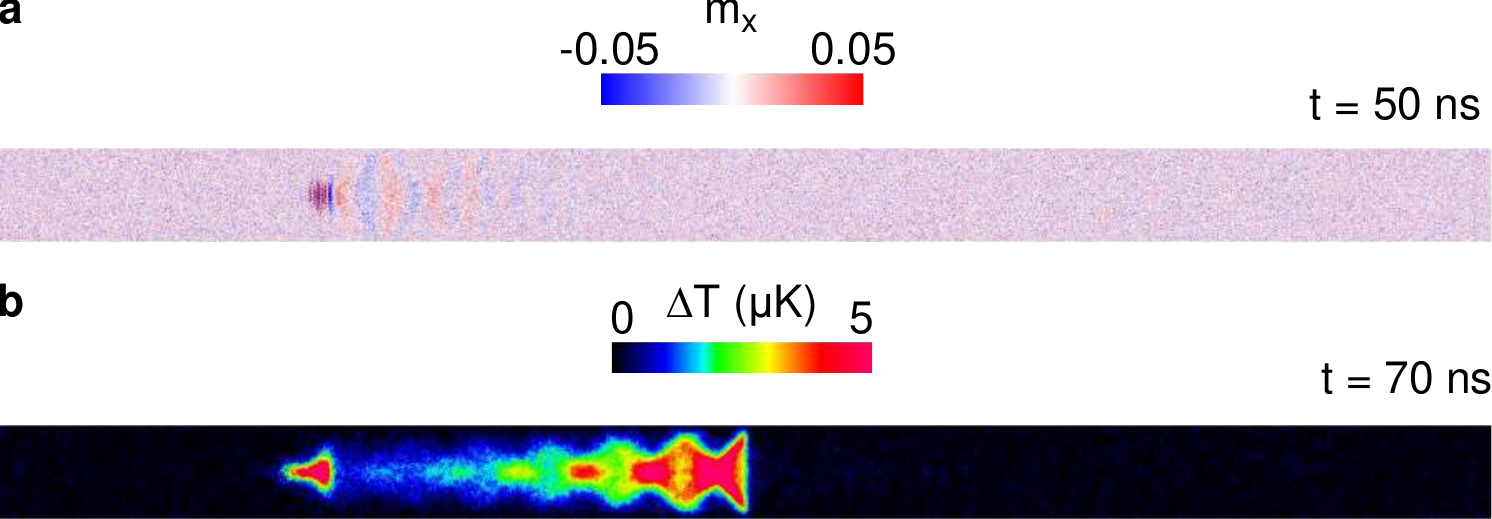}
\end{figure}

\vspace{10ex}
\textbf{Figure S2: Spin wave energy focusing in a reduced size sample. a}, Snapshot of the magnetic configuration 50 ns after the application of a 5 ns-long microwave pulse with frequency 7 GHz. \textbf{b}, Temperature distribution after 70 ns.

\section{Versatility}

One of the main qualities that define the success and viability of a new technique or device is its scalability, that is, the feasibility to reduce the system size without losing its fundamental properties. In order to check if our method of energy control and focusing is scalable, we performed the same simulations on a system 100 times smaller in each direction ($256 \,\mathrm{\mu m} \times 16 \,\mathrm{\mu m} \times 320 \,\mathrm{nm}$), and exciting with an appropriately reduced-size antenna. We found out that the spin wave propagation properties and the energy focusing phenomena remain invariable. However, thermal agitation starts to play a relevant role in smaller systems, and spin-waves may be masked by the random thermal noise. Fig. S2 shows spin wave propagation and energy focusing for a microwave injected current of frequency $7\,\mathrm{GHz}$ and amplitude $1.25\,\mathrm{mA}$. Note that, while this current is much smaller than in the larger system, the magnetic field created by the antenna on the smaller waveguide surface is almost 7 times higher, in order to be able to overcome thermal fluctuations.

Furthermore, exchange plays an important role in smaller systems, and exchange spin waves may be excited in addition to surface waves. In particular, we observe that part of the energy is absorbed in the region where MSSW propagation is forbidden (see Fig. S2b). We attribute it to the excitation of an exchange spin wave at the accumulation region. Exchange spin waves do not have an upper cutoff frequency, and are the only spin waves that can propagate above the MSSW frequency band. The wavelength of the wave obtained from the simulations ($\approx 400\,\mathrm{nm}$) is also compatible with the wavelength obtained from the exchange spin waves dispersion relation at 7 GHz ($\lambda\approx 240\,\mathrm{nm}$). Nevertheless, our computational mesh is not fine enough to accurately describe exchange effects in very large systems, and such study is beyond the scope of the present work. We conclude thus that our system, barring technological limits, is down-scalable at least up to a size of a few micrometers.

\section{Scalability}

While in the main text only magnetostatic surface spin waves are discussed, energy focusing control by magnetic fields is also applicable to magnetostatic backward volume waves (MBVW). MBVW are excited at lower frequencies than MSSW, and their main characteristic is that their wave vector $\vec{k}$ is parallel to the local magnetization. We now magnetize our system along its longitudinal axis, and excite MBVW with a microwave excitation of frequency $5\,\mathrm{GHz}$ with a gaussian profile of width $50 \,\mathrm{\mu m}$. Unlike MSSW, MBVW's wavelength decreases with frequency, so instead of decreasing the external field we have to increase it to observe energy dissipation. We used the same profile from equation S1, with $c=5.625\,\mathrm{mm}$, $w=0.5\,\mathrm{mm}$ and $B_1=-85\,\mathrm{mT}$. Results are shown in Fig. S3.

Although we observe the same energy focusing phenomena, the efficiency drops drastically when considering magnetostatic backwards volume waves. This is due to several reasons. First, MBVW do not propagate unidirectionally, so half of the energy is lost in the wave propagating in the opposite direction. Second, the dispersion relation for MBVW is not as abrupt at the cutoff frequency as in the case of MSSW, so a higher field variation magnitude $B_1$ is needed to achieve the same focusing efficiency. In addition to this, the field variation has opposite sign to the one for MSSW, which renders the proposed experimental realization ineffective for MBVW. Nevertheless, despite the obvious advantages to using surface waves, MBVW may have applications when the geometric or the technological limitations impede the excitation of MSSW.

\begin{figure}[h]
	\includegraphics[width=0.6\linewidth]{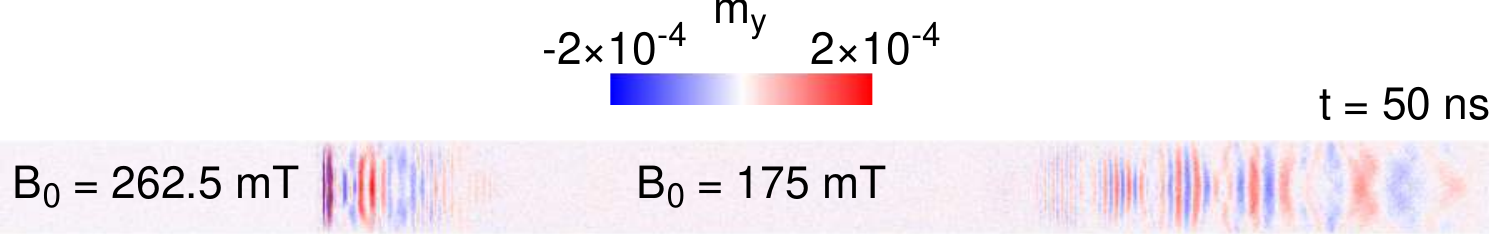}
	
\end{figure}
\textbf{Figure S3: Magnetic field control of MBVW energy focusing.} Snapshot of the magnetic configuration 50 ns after the application of a 5 ns-long microwave pulse with frequency 5 GHz.

\end{document}